# Investigation by STEM-EELS of helium density in nanobubbles formed in aged palladium tritides

B. Evin[a], E. Leroy[b], M. Segard[a], V. Paul-Boncour[b], S. Challet[a], A. Fabre[a], M. Latroche[b]

[a] CEA Valduc, F-21120, Is-sur-Tille, France

[b] Univ Paris-Est Creteil, CNRS, ICMPE (UMR 7182), 2 rue Henri Dunant, F-94320 Thiais, France

E-mail adresses :

berengere.evin@icmpe.cnrs.fr, leroy@icmpe.cnrs.fr, mathieu.segard@cea.fr, paulbon@glvt-cnrs.fr, sylvain.challet@cea.fr, arnaud.fabre@cea.fr, latroche@icmpe.cnrs.fr

Corresponding author :

Bérengère Evin (berengere.evin@icmpe.cnrs.fr)

Phone number : (+33) 149781214

Postal adress : ICMPE-CNRS, 2-4 rue Henri Dunant, 94320 Thiais, France

**ABSTRACT**

$^3$He nanobubbles created by radioactive decay of tritium in palladium tritide are investigated after several years of aging. Scanning Transmission Electron Microscopy – Electron Energy-Loss Spectroscopy (STEM-EELS) has been used to measure helium density from the helium K-edge around 23 eV. Helium densities were found between 20 and 140 (±30) He/nm$^3$ and

the corresponding nanobubble pressures range between 0.1 and 3 (±0.2) GPa. Measuring helium density and mapping He atoms by STEM-EELS enables to differentiate bubbles from empty cavities in the palladium tritide matrix.

## 1. INTRODUCTION

Tritium is of notable interest due to its important role in nuclear industry. Before use, this gas must be safely handled and stored. Hydrogen (or deuterium) gas storage is commonly done by using pressurized tanks but tritium, as a radioactive element, is less compatible with high pressure storage. As a result, reversible solid-state storage by forming metallic tritides (hydrides) at low equilibrium pressure is a suitable solution to store tritium chemically and ensure safety.

Palladium and its alloys are frequently used as tritium storage materials because of their fast kinetics and strong resistance to oxidation [1]. Tritium spontaneously decays into helium-3 ($^3$He) with a half-life decay $t_{1/2} \approx 12.3$ years. Contrary to tritium, $^3$He solubility in Pd-based alloys is very small. Consequently $^3$He nanobubbles nucleate within the Pd matrix by a self-trapping process [2]. The formation of $^3$He nanobubbles comes with the creation of defects like auto-interstitial metal atoms, dislocation loops and dislocations that modify the structural, microstructural, mechanical and thermodynamic properties of the metal tritides [3],[4],[5] [6] [7]. A recent review on implanted $^3$He bubbles is available in reference [8]. This phenomenon is commonly named the "aging" process.

Palladium-based material aging has been studied at different scales using complementary methods. Correlations between macroscopic mechanical studies and Transmission Electron Microscopy (TEM) measurements show that the Young modulus growth stops after one week whereas $^3$He bubble density stabilizes after the first month of aging [9]. The knowledge of the distribution of the $^3$He nanobubbles in the aged material is of noteworthy interest to better understand the modification of the metal tritides properties. The material modification and evolution during aging is calculated using a model that requires experimental input data [10].

After 1 to 3 months of aging in Pd, TEM studies have revealed that $^3$He bubbles have a mean diameter of 1 nm and are uniformly distributed [11],[12]. For longer aging time (8 months to several years), the $^3$He bubble diameters vary between 1 and 25 nm with a distribution centered around ~2 nm [13] [14]. The mean bubble size does not evolve too much after few months of aging.

Beside bubble size and density, $^3$He pressure inside the nanobubbles is also an important parameter for modelling. Nuclear Magnetic Resonance (NMR) [15] or Scanning Transmission Electron Microscopy – Electron Energy-Loss Spectroscopy (STEM-EELS) [16] data can be used to calculate the $^3$He bubbles pressures. NMR gives access to the average pressure of the whole bubble population by following the rigid-fluid transition of $^3$He, for which the temperature depends on the pressure [17]. Complementarily, STEM-EELS allows for the local probe the pressure inside an individual bubble by following the 1s → 2p transition of $^3$He. In the present work, this last technique is used to characterize Pd powders with different aging time and to determine the $^3$He pressure in the nanobubbles as a function of their size.

STEM-EELS allows the combination of both acquisitions imaging and spectroscopy modes. STEM is a structural measurement that enables to observe bubbles and determine their size whereas EELS is a quantitative way giving information on the He density. Indeed, He free atoms exhibit a 1s → 2p transition around 21.2 eV [18]. Walsh *et al*. proposed a method to calculate He density in a nanobubble based on the blueshift of this transition [16]. He density inside the cavities can also be determined by the intensity of the transition peak.

He density in bubbles was measured using this procedure for several materials, mostly obtained by He implantation. David *et al*. evidenced the impact of residence time on He density in nanobubbles implanted in semiconductor (silicon and germanium) matrix. Bubbles between 5 and 25 nm in diameter show densities from 80 to 110 He/nm$^3$ corresponding to pressures between 2 and 6 GPa [19]. Lacroix *et al.* used spatially resolved EELS to investigate He -filled pores in cobalt created by magnetron sputtering with nanopore sizes ranging from 4 to 20 nm [20]. He densities range between 10 and 100 He/nm$^3$ for pressures of 0.05 to 5.0 GPa. The authors also observed the weaker He transition 1s → 3p around 25 eV [20]. Blackmur *et al.* mapped the association of hydrogen close to implanted He bubbles surface (between 1 and 10 nm diameter) in zirconium. In addition, halos were seen at 13.5 eV in the EELS data around the bubbles and were attributed to hydrogen [21].

Taverna et al. used STEM-EELS to study He nanobubbles (diameter from 2 to 25 nm) within bulk Pd$_{90}$Pt$_{10}$ alloy aged for 8 months under tritium [13]. He densities range from 15 to 35 He/nm$^3$ corresponding to pressure between 0.1 and 0.3 GPa. He peak shift was also studied, and a linear relationship between blueshift and He density was obtained. Overall, for all these nanobubbles the order of magnitude expected for the pressure is close to GPa. All the quoted works are consistent with the theoretical trend: the pressure decreases with increasing the bubble diameter (Laplace's law : $P=2\gamma/r$, with $P$ the pressure, $\gamma$ the surface tension and $r$ the bubble radius).

Complementary to these previous works, the present investigation focusses on the identification and characterization of $^3$He nanobubbles obtained by tritium radioactive decay in Pd powders with longer aging times (several years). Spatially resolved Electron Energy Loss Spectroscopy has been used to measure and map $^3$He density and the corresponding pressure for two samples that are several years old.

## 2. EXPERIMENTAL METHODS

### 2. 1. Samples preparation

Palladium powder samples (mean aggregates size about 14 µm, mean particle size between 0.2 and 1 µm) were cleaned up by deuterium absorption–desorption cycles before tritium loading, and were aged at room temperature. Pd powders were aged during 6 to 8 years under tritium and have been maintained at a constant T/Pd ratio during aging time The decontamination of Pd tritide powders is realized by isotopic exchange with deuterium in β-phase.

To characterize the aging of the materials, the ratio between $^3$He and metal atoms (He/M) was used. It is calculated using the tritide stoichiometry and the aging time. In the present work, measurements were performed on two samples with He/Pd ratios of 0.23 and 0.27 as they contain large $^3$He bubbles [14]. Their close aging time implies that no large difference is expected between these two samples, but it allows checking the reproducibility of the results from one sample to another.

The powders were embedded in an epoxy resin. Samples were cut using ultramicrotomy at room temperature with a diamond knife and deposited on a 400 mesh copper grid. Before TEM observations, the sample surface was cleaned in an $O_2/H_2$ plasma in order to remove the carbonated contamination.

## 2. 2. STEM-EELS characterization

The experiments were conducted in a FEI TECNAI F20 G2 microscope operated at 200 kV. The STEM-EELS acquisitions were recorded through a GATAN GIF 2001 with the following experimental parameters: 1024 channels, energy dispersion of 0.1 eV/channel, integration time of 0.1 s, pixel size ranging from 1 to 2 nm, energy resolution of 1.4 eV, convergence and collection angles of 17 mrad and 5.86 mrad respectively, beam current of the electron probe of 9.87 nA and probe size of 1 nm. The acquisition time was optimized to maximize the $^3$He K-edge signal, avoid sample drifting and avoid $^3$He detrapping from nanobubbles. Simultaneously, High Angle Annular Dark Field (HAADF) images of the STEM-EELS investigated area were recorded. The mean free path is around 100 nm.

Moreover, to reduce the probability of $^3$He detrapping and enhance the sample stabilization under the electron beam, all the experiments were conducted at temperatures ranging from 101 K to 141 K with a GATAN double tilt nitrogen cooling sample holder model 636. Liquid nitrogen cooling was efficient in reducing $^3$He detrapping and in slowing carbon contamination. These observations were made after conducting first STEM-EELS attempts at room temperature. Measurements at 101K showed no helium detrapping after twelve acquisitions on the same bubble. This phenomenon was also reported in reference [22] and linked to long time scale relaxation phenomenon.

## 3. RESULTS AND DISCUSSION

### 3. 1. Data analysis

#### 3. 1. 1. TEM Imaging

In this study, « cavity » refers to any cavity filled or not with $^3$He whereas « bubble » stands for $^3$He-filled cavity by opposition to « empty cavity ». TEM imaging was performed to measure the cavity number and diameters for several palladium tritide grains. Under-focused (-1 µm) images reveal cavities as white dots and over-focused (+1 µm) images as black dots as seen in Figure 1. The cavity surface density is measured using the maximum search of the ImageJ software and is combined with the sample thickness measured by EELS (see Supplementary Materials 5 and 6) to obtained the cavity volume density in the material (in cavity/m$^3$). The cavity diameter is measured using the Feret's diameter. For each sample, at least 3000 cavities are processed to build the statistics on cavity diameter and density.

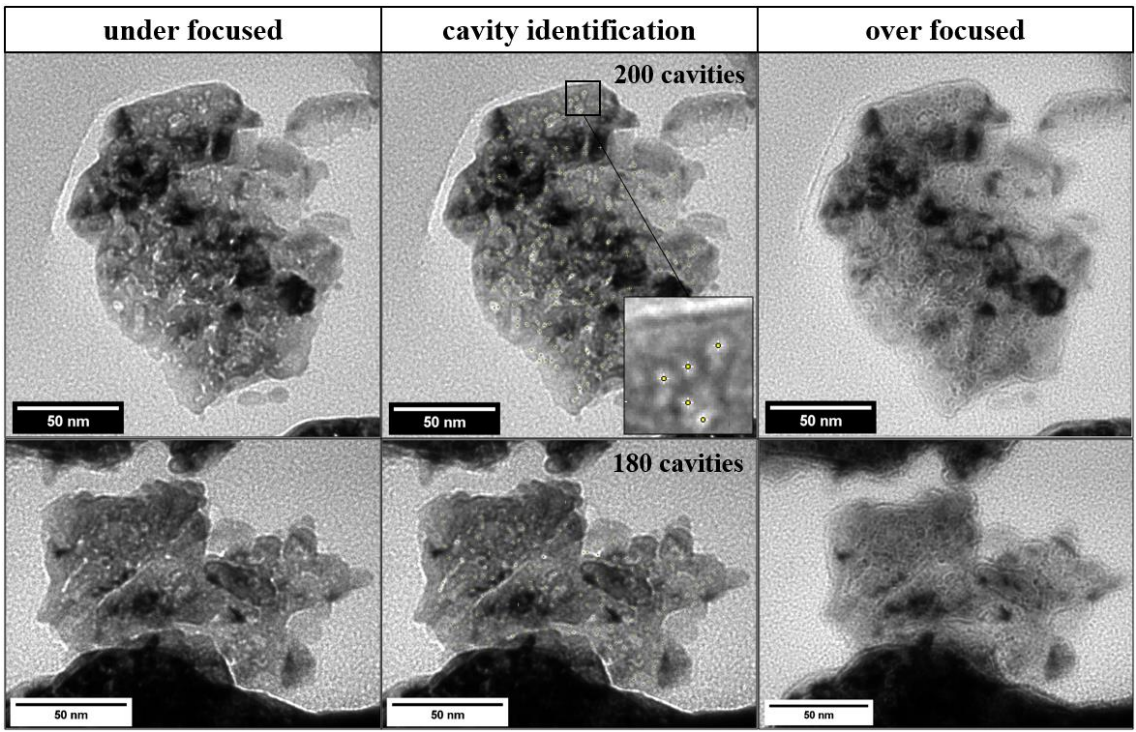

**Figure 1 - Under (left) and over (right) focused TEM images acquired on two aged palladium tritide grains with He/Pd=0.23.**

In both samples (He/Pd = 0.27 and He/Pd = 0.23), the cavity diameters range between 2 and 20 nm. The cavity size exhibits a log-normal distribution centered around 2.5 nm and displays a mean value between 2.6±0.3 and 2.8±0.3 nm (see Supplementary Materials 1). These cavity diameters are slightly larger than those measured by Thiébaut *et al.* [11] (0.8 to 1 nm) for Pd alloys aged several months compared to several years in this work. Cavity density ranges from $4.10^{23}$ to $1.10^{24}$ cavity/m$^3$ in the present samples. The cavity density obtained by Thiébaut *et al.* was around $10^{25}$ bubble/m$^3$, slightly higher but not far from the current results. The discrepancy can be explained by the uncertainty in the determination of the sample thickness in reference [11]. Indeed, the cutting thickness was used to calculate the cavity volume density. In the present study, the sample local thickness was measured by EELS spectroscopy and used to calculate the cavity volume density. The cutting thickness can be slightly different than the real local thickness and its value will directly modify the cavity volume density.

### 3. 1. 2. Helium nanobubble EELS spectrum

In HAADF images, cavities appear as black dots in a light grey palladium matrix. From extracted EELS spectra, the presence of $^3$He appears as a sharp peak around 23 eV attributed to the $^3$He K-edge (Figure 2). Besides $^3$He, Pd contribution appears as a broad plasmon centered around 25 eV and spreading between 15 and 40 eV. These different shapes between the EELS spectra of $^3$He and Pd allow clear distinction between $^3$He nanobubbles from the Pd matrix (Figure 2 and 3).

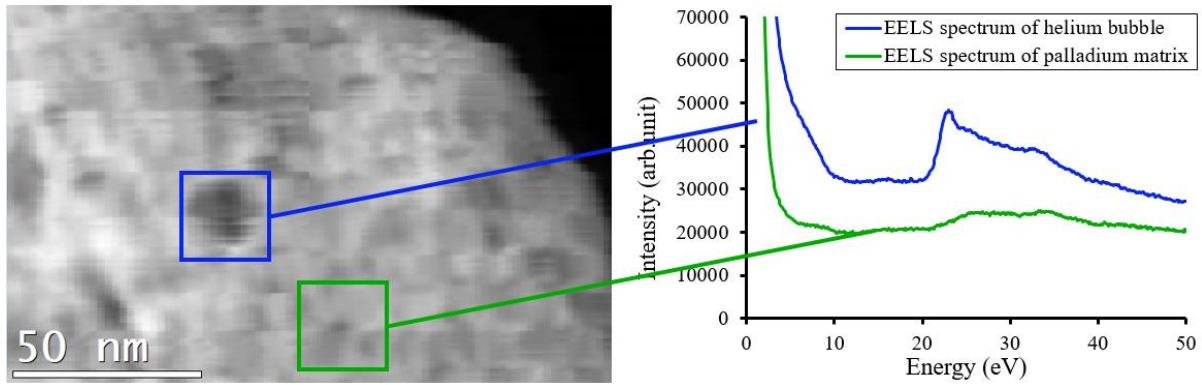

**Figure 2 - HAADF image (left), extracted EELS spectrum of helium bubble and Pd matrix (top blue), extracted spectrum of Pd matrix (bottom green) acquired on aged palladium tritide grains with He/Pd=0.27.**

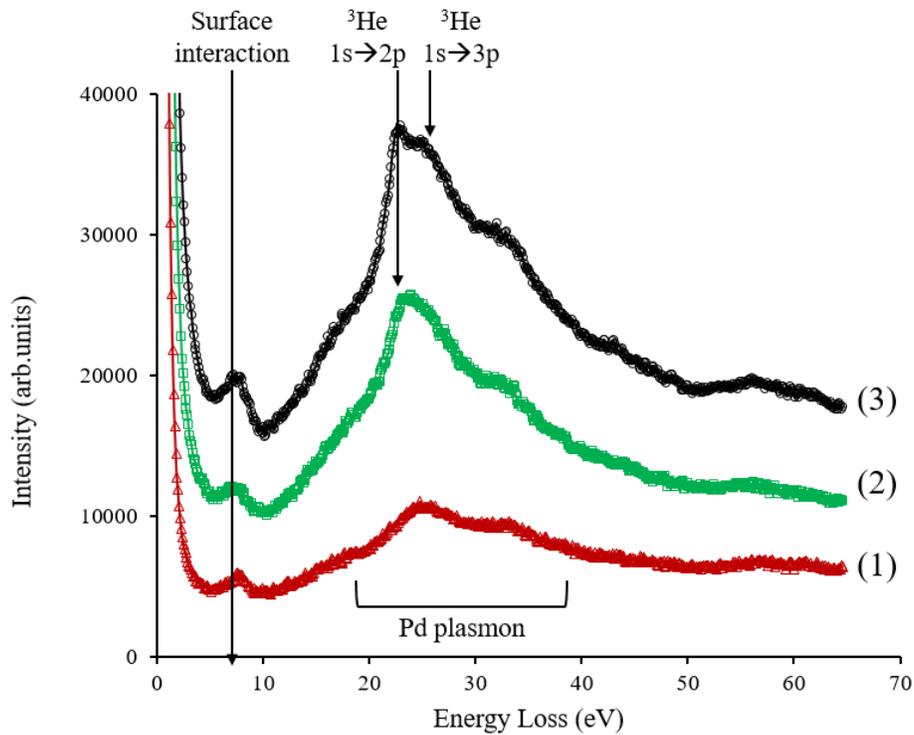

**Figure 3 - EELS spectra of Pd matrix (1), Pd matrix + He bubble with 1s→2p transition (2), Pd matrix + He bubble with 1s→2p transition and 1s→3p transition (3) acquired on aged palladium tritide grains with He/Pd=0.23**

Three different EELS spectra are presented in Figure 3: the red one is extracted from the Pd matrix, the green and black ones arise from $^3$He nanobubbles. For the black spectrum, beside the main 1s → 2p He-K edge transition at 23 eV, a second peak close to 25 eV is observed and attributed to the 1s→3p He-K edge transition. This weaker transition has also been reported by Blackmur et al. [21]. However, this side peak is not always observed in our measurements.

In addition to the main peaks in Fig. 3, a small peak is also seen at 7 eV. It is commonly attributed to surface interactions between the $^3$He bubbles and the matrix [19],[21]. This peak is usually better seen with EELS spectra extracted close to the bubble edge. Since $^3$He nanobubbles are included in a Pd matrix, the resulting EELS spectrum is the sum of $^3$He and Pd energy-losses, $^3$He K-edge and Pd plasmon are then overlapping.

### 3. 1. 3. Datacube treatment

The first step to process EELS spectra and to calculate $^3$He density consists in the extraction of the $^3$He contribution by removing the palladium plasmon signal. Several methods reported in the literature have been tested : spatial difference [21], Principal Component Analysis (PCA, MVA…) and Gaussian adjustment [23]. The latter one was found to be the best method to remove the Pd plasmonic contribution in our case (Figure 4).

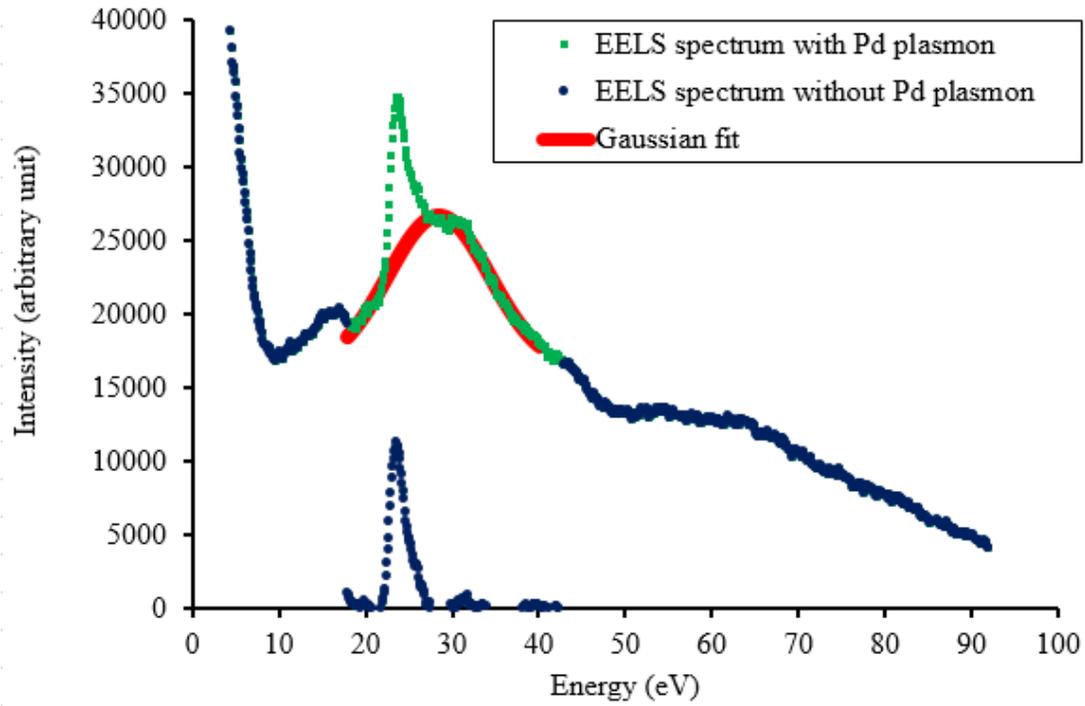

**Figure 4 -  EELS spectrum before (green square □) and after (blue circle ○) subtraction of the Pd plasmonic contribution using Gaussian adjustment (red curve).**

The refinement was done with a Gaussian like function, first by selecting points before (15-20 eV) and after (30-45 eV) the ³He K-edge, and then extrapolating to the entire range between 15 and 45 eV. This method correctly extracts the ³He peak as shown in Figure 4.

### 3. 1. 4. Atomic density

The ³He density inside the bubbles was calculated using the helium 1s→2p transition and the following equation (1):

$$n_{He} = \frac{I_{He}}{I_{ZLP} \sigma_{He} h} \qquad (1)$$

$n_{He}$ represents the helium volume density (in He/nm$^3$), $I_{He}$ is the integrated signal of helium, $I_{ZLP}$ the integrated intensity of the elastic peak (zero-loss peak), $\sigma_{He}$ the cross-section (in nm²) and $h$ the cavity thickness crossed by the electron beam (nm).

The convergence and collection angles were measured and are respectively 17.0 and 5.86 mrad. The cross-section $\sigma_{He} = 5.9.10^{-6}$ nm² was calculated using the Sigmak3 program [24]. Since the sample mostly contains Pd, the absolute thickness of the sample is calculated using the atomic number of Pd. However, this does not give the He thickness, needed for the volume density calculation. An approximate He thickness can be obtained using the projected diameter of the bubble.

The uncertainty associated with the bubble diameter is taken equal to the spatial resolution, that is 1 nm. The cartography of the surface density is the result of calculation and post-processing; its uncertainty is complex to evaluate. The penalizing choice of assimilating it to the surface density which was sometimes observed on the areas of Pd matrix, where it should be zero but due to treatment can reach 100 He/nm² was made (see Supplementary Materials 3 and 4). These calculations lead to an average uncertainty of 30 He/nm$^3$ that will be used in this study. This leads to a pressure uncertainty of 0.2 GPa.

### 3. 1. 5. Helium density mapping

Once the He peak is isolated in EELS spectra, He volume density is determined using equation (1). This equation is implemented in a homemade Digital Micrograph script to map $^3$He surface density on the surveyed area of the sample as seen in Figure 5. STEM-EELS data in Figure 5 were recorded at 101 K.

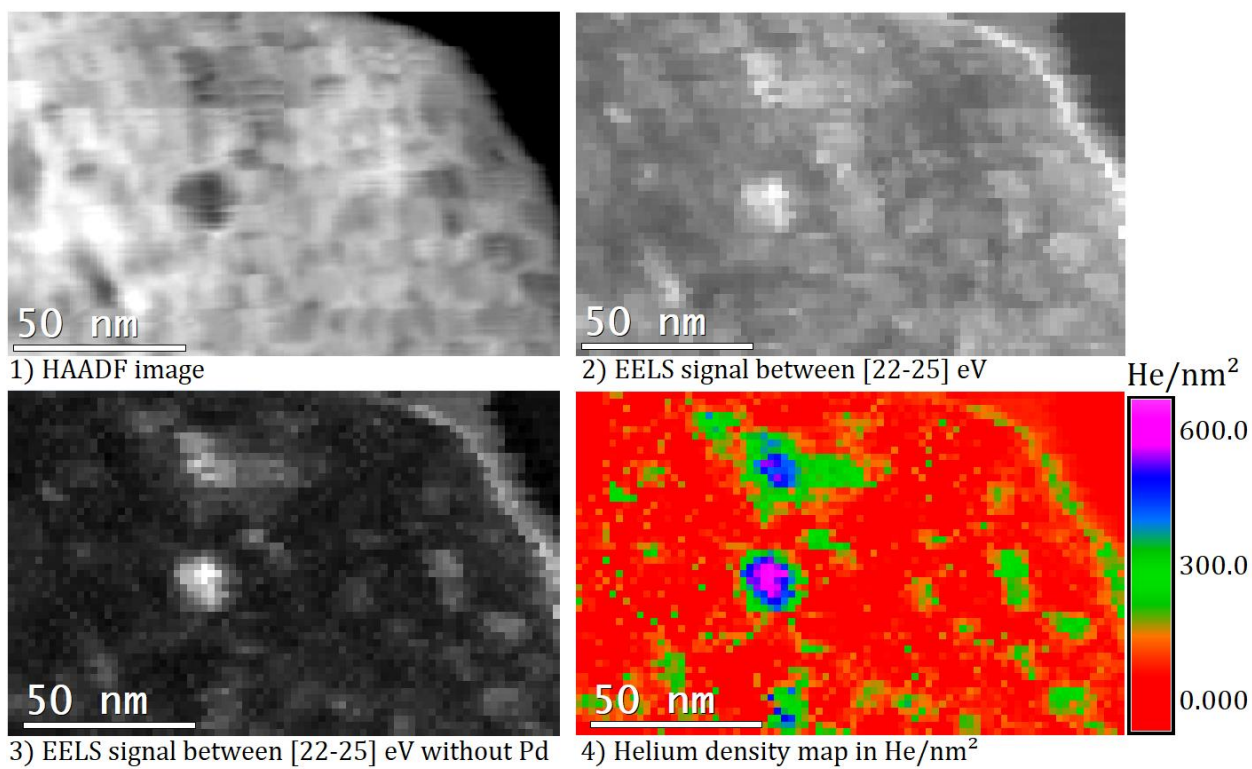

**Figure 5 - 1) HAADF image (top-left) 2) EELS signal between 22 and 25 eV (top-right) 3) EELS signal between 22 and 25 eV after data treatment (bottom-left) 4) Helium surface density map calculated in He/nm² (bottom-right) for aged palladium tritide grains with He/Pd=0.27. All images were recorded at 101 K.**

The white pixels in Figure 5-2 represent the high intensity EELS signal between 22 and 25 eV and are consistent with the black areas shown in HAADF image, Figure 5-1 allowing the identification of $^3$He bubbles. Subtraction of the Pd plasmon contribution (Figure 5-3) enhances the contrast of the EELS signal between bubbles and Pd. Finally, the $^3$He mapping in He/nm² shown in Figure 5-4 enables the easy identification of areas of high $^3$He concentrations, i.e. $^3$He nanobubbles. The ratio between the helium density mapped in He/nm² and the diameter of the bubble allows determination of the helium density in He/nm$^3$. It is also worthwhile to note the presence of black dots in the HAADF image that do not correspond to $^3$He bubbles and are therefore interpreted as empty cavities. A $^3$He surface density (in He/nm²) gradient is noticeable from the center to the edge of the main bubbles.

### 3. 2. Results

### 3. 2. 1. Multiple nanobubbles

The STEM-EELS acquisition was done close to one large bubble (⌀ 11 nm) at 101 K. The HAADF image (Figure 6-left) reveals that the bubble might be the sum of two overlapping bubbles as two darker spots are noticeable in the bubble.

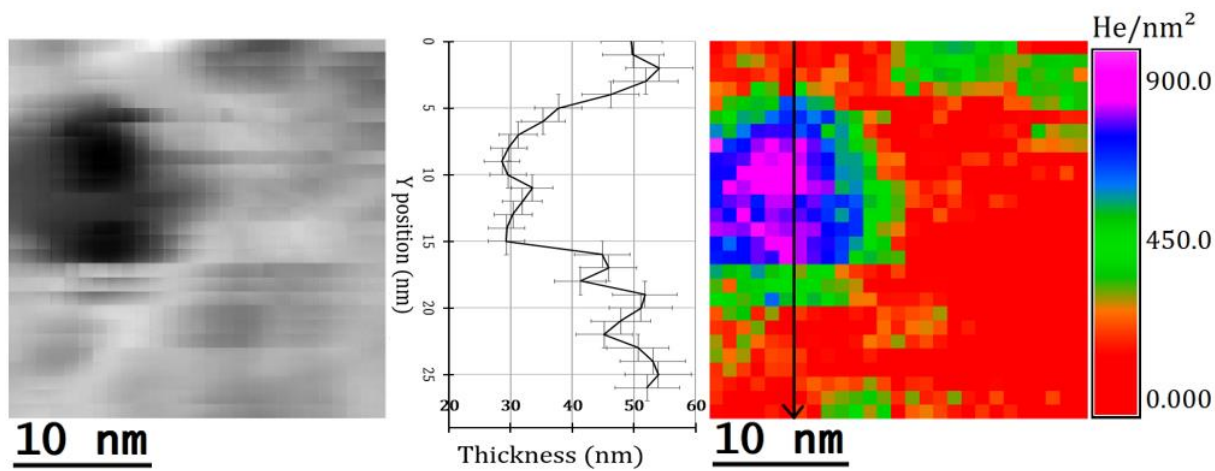

**Figure 6 - HAADF image (left), thickness profile (center) and helium surface density map in He/nm² (right) of helium bubbles for aged palladium grains with He/Pd=0.27. The thickness profile was measured along the black arrow shown on the right.**

The $^3$He surface density (Figure 6-right) is higher at the center of the bubble (around 900 He/nm²) than at the edges (500 He/nm²). The mean density of the entire bubble is around 600 He/nm². The ratio between the $^3$He projected density in He/nm² and the projected diameter of the bubble allows determination of the $^3$He volume density in He/nm$^3$. The mean density of the whole bubble is then 60±30 He/nm$^3$ which corresponds to a pressure of 1.1±0.2 GPa. For the two smaller bubbles that seem included in the large one, the densities are 140±30 He/nm$^3$ (top bubble) and 200±30 He/nm$^3$ (bottom bubble). Since bubble pressure (or density) is expected to increase with decreasing diameter, this last result is not surprising. This inverse proportional relation between pressure and radius following Laplace's law was reported in reference [13].

A thickness profile of the Pd sample (Figure 6-center) was extracted along the black arrow shown on the $^3$He density map (Figure 6-right). The sharp decrease of the thickness from 50 nm to 30 nm confirms the presence of a bubble. Since the projected diameter of the whole bubble is 11 nm, the 20 nm thickness drop favors the assumption of overlapped bubbles. Moreover, the slight increase from 30 to 35 nm at the center of the bubble supports the hypothesis of two overlapped bubbles in the thickness of the sample. Nevertheless, EELS spectroscopy gives only a 2D projected information from a 3D sample, it is then difficult to discriminate between one or two bubbles. This issue could be solved by tilting the sample to obtain spatial information by tomography, but this is out of the scope of this paper.

### 3. 2. 2. Empty cavities

At the beginning of this section, the difference between empty cavity and bubble was made. Figure 7 illustrates this case for a sample with He/Pd = 0.27, in which the HAADF image shows five dark circular spots.

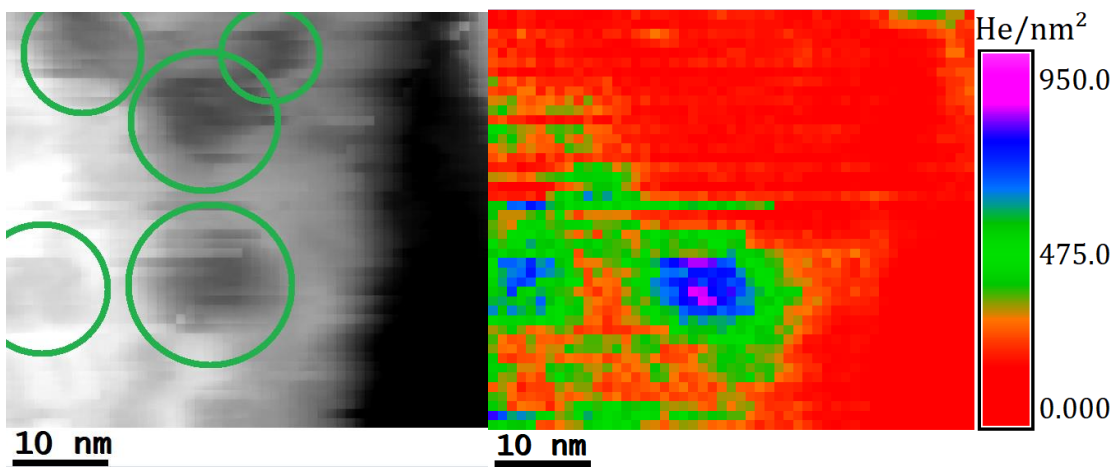

**Figure 7 - HAADF image (left) with cavities circled in green and helium density map in He/nm² (right) of helium bubble and empty cavity for aged palladium tritide grains with He/Pd=0.27.**

The HAADF image (Figure 7 left) shows five cavities of similar morphology seen as large black circular dots circled in green. HAADF intensity profile is available in Supplementary Materials 7. However, the $^3$He density map (Figure 7 right) only reveals $^3$He in two cavities. The other three cavities are either empty or have too low pressure to observe He. The $^3$He density of the left bubble is 75±30 He/nm³ corresponding to 1.7±0.2 GPa, and the density in the right bubble is 50±30 He/nm³ corresponding to a pressure of 0.7±0.2 GPa.

$^3$He might have left the cavities during the EELS measurement due to interaction with the electron beam. This phenomenon was quantified by David et al. and was surprisingly fast in their case [19] [22]. This emptiness can also be explained by the closeness between the cavity and the Pd surface or by some $^3$He departure during sample preparation. Pre-existing cavities have been observed in hydrogenated Pd

powders not aged under tritium (see Supplementary Materials 2). This might explain the origin of some empty cavities.

Distinction between bubble and cavity illustrates the strength of EELS compared to other analytical techniques. Indeed, TEM and SANS allow the identification of cavities, whereas NMR can measure the mean He pressure in bubbles but does not detect cavities.

### 3. 2. 3. Radial evolution of helium density

The radial evolution of $^3$He density from the center to the edge of the bubbles has been studied for the largest ones in the samples.

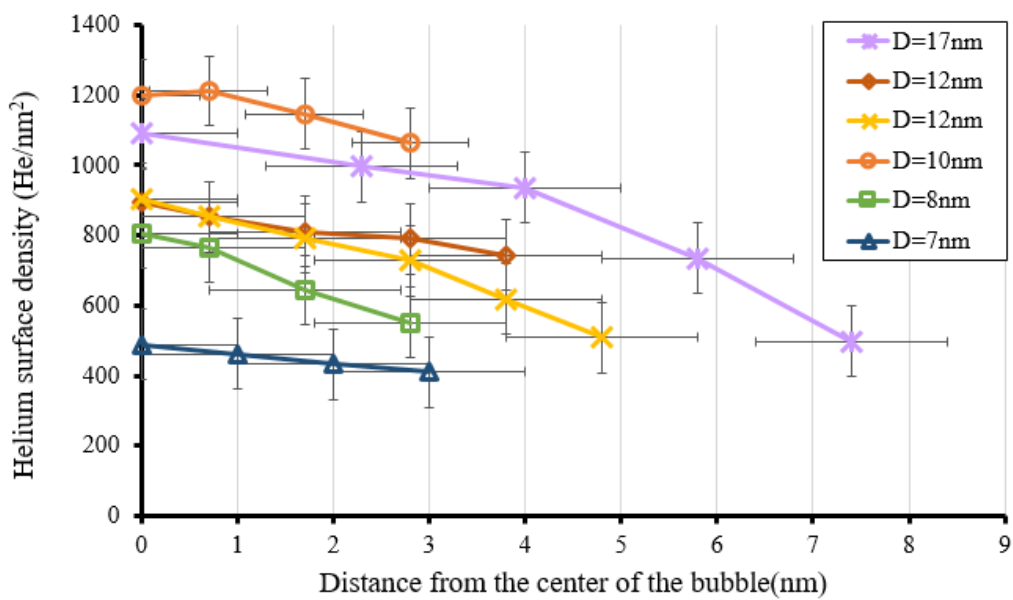

**Figure 8 - Radial evolution of helium density in He/nm$^3$ in aged palladium tritide grains with He/Pd=0.23 and 0.27.**

For all bubbles, $^3$He surface density decreases slightly as the value is taken far from the center of the bubble. This gradient of density in He/nm² is seen on bubbles taken from Figure 6 and

7. This tendency is related to the fact that the beam encounters less ³He atoms at the edge of the bubbles, however this effect remains small except for the ⌀ 17nm bubble. This small effect supports the assimilation of ³He thickness with the cavity diameter.

### 3. 2. 4. Density-Pressure-Diameter relationship

From all studied samples, the ³He density range was established between 20 and 140±30 He/nm³ with an average density of 70±30 He/nm³. From ³He density, pressure was calculated using semi-empirical equation of state (EOS) of helium-4 assuming that there is no isotopic effect between ³He and ⁴He. The analytical expression of the EOS was extracted from references [19] and [25] for high and low densities (under/over 140 He/nm³ respectively).

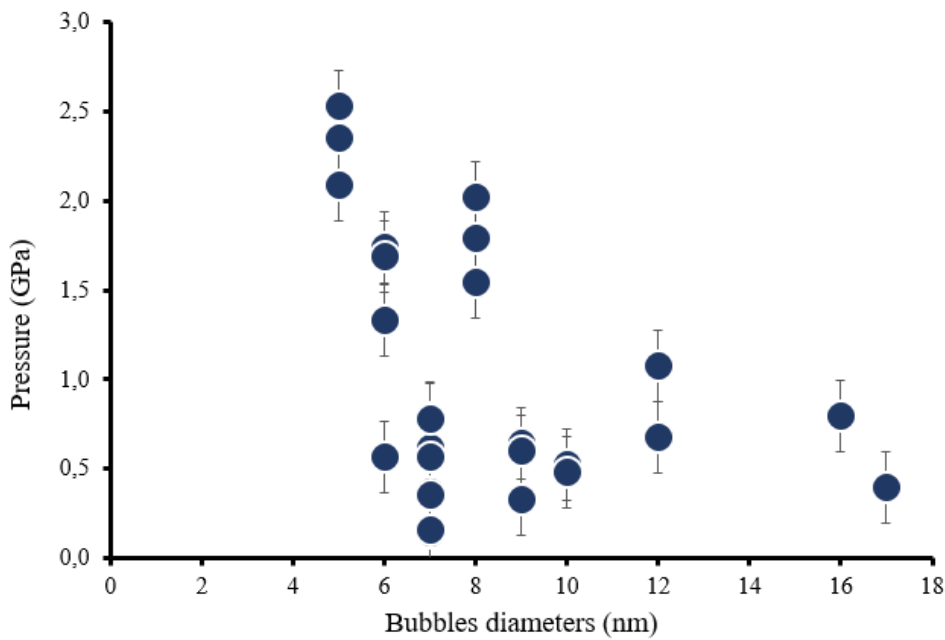

**Figure 9 - Helium pressure as a function of bubble diameters in aged palladium tritide grains with He/Pd = 0.23 and 0.27.**

As seen in Figure 9, the $^3$He pressure ranges from 0.1 to 3 ± 0.2 GPa for bubbles ranging from 17 to 5 nm, the average pressure being 1.1±0.2 GPa. The pressure tends to decrease when the bubble diameter increases, as predicted by the Laplace's law. The pressure values found in the $^3$He bubbles agree with previous works. Taverna *et al.* found pressures ranging from 0.1 to 0.3 GPa for bubbles of 17 to 5 nm, respectively in bulk Pd-Pt alloy [13]. David *et al.* found pressures ranging from 2 to 6 GPa for bubbles from 25 to 6 nm in silicon [19].

The bubble pressure is a key parameter calculated by modelling the bubble growth mechanism in metallic tritide, based on continuum mechanics [10]. The order of magnitude of the pressure provided by the model is a few GPa, in good agreement with the pressures determined by STEM-EELS. The experimental bubble pressure determinations enhance the confidence in the theoretical model.

Mean helium pressure estimated by NMR in aged Pd sample was estimated around several GPa for a transition temperature of 250 K [15] [26] [27]. The combination of these previous NMR results and this EELS study strengthens the confidence placed in the theoretical results provided by the mechanical model (bubble pressure of several GPa). Further works will be performed with samples having similar He/Pd rate and measurements will be done at liquid He temperature to compare EELS and NMR techniques. In addition, the difference between values from EELS and NMR can be explained by the size of the bubbles. By EELS analyses, a focus is made on the largest bubbles (∅ > 5 nm) whereas NMR gives an average value including smaller bubbles.

Since mean diameter of the samples is 2.6 nm, pressure obtained by EELS on large bubbles may be under the average value of the samples.

## 4. CONCLUSION

$^3$He nanobubbles created by radioactive decay of tritium in palladium tritide powders were successfully characterized by STEM-EELS. HAADF imaging led to the localization of the cavities within the Pd matrix, whereas EELS spectroscopy allowed the identification and quantification of $^3$He bubbles.

Data treatment (data post-processing and use of a homemade script) of the EELS spectra allows the extraction of the $^3$He contribution from the Pd signal and the mapping of the $^3$He density in the surveyed areas. The $^3$He densities, ranging between 20 and 140±30 He/nm$^3$, agree with previous studies. For $^3$He nanobubbles from 5 to 17 nm diameter, the $^3$He pressures decrease from 3 to 0.1±0.2 GPa.

Those pressures determinations enhance the confidence in the theoretical mechanical model of helium bubble growth that provides bubble pressures of a few GPa.

Mapping $^3$He density in He/nm² reveals that $^3$He surface density decreases from the center to the edge of the bubbles. This tendency is consistent with the fact that the electron beam encounters more $^3$He atoms at the center of the bubble.

The STEM-EELS measurements enable to better characterize the $^3$He bubbles observed by TEM. By measuring $^3$He density and pressure, it is possible to discriminate between bubbles and empty cavities. Some interactions between the electron beam and $^3$He atoms were observed, which can induce $^3$He diffusion and detrapping.

Further investigations with STEM-EELS on $^3$He bubbles will focus on spatial resolution and thickness evaluation that will be enhanced by calculating the number of missing metal atoms [13],[19]. Moreover, samples with different aging times will be studied. To improve further the knowledge on $^3$He nanobubbles and Pd tritide aging, next work will focus on 3D characterizations using electron tomography.

**FUNDINGS**

This work was supported by the France's Atomic Energy Commission (C.E.A.) and by the French National Center for Scientific Research (C.N.R.S.).